\documentclass[useAMS,usenatbib]{mn2e}

\usepackage{graphicx}
\usepackage{txfonts}
\usepackage{enumerate}
\usepackage{xcolor} 

\newcommand{\sbs}{SBS\,0335$-$052}
\newcommand{\izw}{I\,Zw\,18}
\newcommand{\aap}{A\&A}
\newcommand{\aaps}{A\&AS}
\newcommand{\apj} {ApJ}
\newcommand{\apjl} {ApJL}
\newcommand{\apjs} {ApJS}
\newcommand{\aj} {AJ}
\newcommand{\araa} {ARA\&A}
\newcommand{\mnras}{MNRAS}
\newcommand{\nat}{Nature}

\newcommand{\gca}{Geochimica et Cosmochimica Acta}
\newcommand{\pasp}{PASP}
\newcommand{\hers}{{\it Herschel}}
\newcommand{\logoh}{12$+$log(O/H)}
\newcommand{\zzsun}{$Z/Z_\odot$}
\newcommand{\zsun}{$Z_\odot$}
\newcommand{\msun}{$M_\odot$}
\newcommand{\lsun}{$L_\odot$}
\newcommand{\msunyr}{$M_\odot$\,yr$^{-1}$}
\newcommand{\msunpc}{$M_\odot$\,pc$^{-2}$}
\newcommand{\htwo}{H$_2$}
\newcommand{\hi}{H{\sc i}}
\newcommand{\hii}{H{\sc ii}}

\newcommand{\meannhtwo}{$\langle\,n_{\rm mol}\,\rangle$}

\newcommand{\cmtwo}{cm$^{-2}$}
\newcommand{\cmthree}{cm$^{-3}$}
\newcommand{\magarc}{mag\,arcsec$^{-2}$}
\newcommand{\dusty}{{\it DUSTY}}
\newcommand{\sigmahi}{$\Sigma_{\rm HI}$}
\newcommand{\sigmahtwo}{$\Sigma_{\rm H2}$}
\newcommand{\mstar}{M$_{\rm star}$}

\title[The dust content of the most metal-poor galaxies]{The dust content of the most metal-poor star-forming galaxies}
\author[Raffaella Schneider, Leslie Hunt and Rosa Valiante]{Raffaella Schneider$^{1}$\thanks{E-mail: raffaella.schneider@oa-roma.inaf.it}, Leslie Hunt$^{2}$ and Rosa Valiante$^{1}$ \\
$^{1}$INAF/Osservatorio Astronomico di Roma, Via di Frascati 33, 00040 Monte Porzio Catone, Italy\\
$^{2}$INAF/Osservatorio Astrofisico di Arcetri, Largo Enrico Fermi 5, 50125 Firenze, Italy }
\begin{document}

\date{draft version 20 November 2015}

\pagerange{\pageref{firstpage}--\pageref{lastpage}} \pubyear{2015}

\maketitle

\label{firstpage}

\begin{abstract}
Although dust content is usually assumed to depend uniquely on metallicity,
recent observations of two extremely metal-poor dwarf galaxies have suggested that 
this may not always be true.
At a similar oxygen abundance of $\sim$3\%\,\zsun, 
the dust-to-gas and dust-to-stellar mass ratios in \sbs\ and \izw\
differ by a factor 40-70 according to including molecular gas or excluding it.
Here we investigate a possible reason for this dramatic difference through models
based on a semi-analytical formulation of chemical evolution including dust.
Results suggest that the greater dust mass in \sbs\ is due to the more efficient grain
growth allowed by the high
density in the cold interstellar medium (ISM), observationally inferred to be almost 20 times 
higher than in \izw.
Our models are able to explain the difference in dust masses, suggesting that efficient
dust formation and dust content in galaxies,  including those with the highest measured redshifts, 
depend sensitively on the ISM density, 
rather than only on metallicity.
\end{abstract}


\section{Introduction}
\label{sec:intro}

The dust content in galaxies is intimately linked to their evolutionary history.
Nevertheless, the mass of the dust in the interstellar medium (ISM), 
and its ratio with the gas mass (dust-to-gas ratio, DGR), are critical parameters
for establishing the evolutionary state of a galaxy.
However, the complex interplay between dust destruction and dust formation mechanisms
\citep[e.g.,][]{dwek98,bianchi07,jones11},
makes it difficult to infer evolutionary trends from dust alone.

Another signature of evolutionary status is a galaxy's metal abundance, 
generally quantified by the nebular abundance of oxygen, O/H.
Although dust grains consist mainly of metals,
grains are not the dominant contributor to the metal budget of galaxies
\citep{peeples14}.
Nevertheless, the metals in dust and the metals in gas are expected to be coupled
through the ISM energy cycle, and
much work has been focused on comparing dust content with metallicity.
Indeed, the DGR and O/H seem to be fairly well correlated both in the Milky Way and in 
nearby star-forming galaxies.
Observations of the DGR compared with the gas-phase metallicity 
suggest approximate linearity between the two
\citep[e.g.,][]{issa90,schmidt93,draine07sings}, a trend which is generally true even 
for gradients of the DGR within galaxies, which tend to follow the radial changes in
metallicity \citep[e.g.,][]{munoz09,magrini11}. 
Since the seminal work by \citet{dwek98}, many galaxy evolution models and models of the ISM
now {\it assume} that dust content is proportional to metallicity
\citep[e.g.,][]{granato00,wolfire03,gnedin09,krumholz09b}. 
In the ISM models, this assumption is directly related to the capability of the gas
to self-shield from intense ultraviolet radiation, and is thus crucial for the formation
of \htwo.

However, new results with \hers\ and ALMA challenge the assumption of a direct (linear) correspondence
of the DGR with metal abundance. 
Based on dust masses calculated with \hers\ data from the Dwarf Galaxy Survey \citep{madden13},
\citet{remy14} found that the DGR is linearly proportional to O/H only to \logoh$\sim$8 (20\%\,\zsun);
at lower metallicities the dependence of DGR is steeper, implying relatively less dust (or more gas)
at lower abundances.
By including new ALMA data at 870\,\micron, \citet[][hereafter H14]{hunt14} derive vastly different DGRs
even at the same metallicity, in particular for the two most metal-poor star-forming galaxies
in the Local Universe, \sbs\ and \izw, both at $\sim$3\%\,\zsun.
As shown in Fig. \ref{fig:dgroh} (adapted from \citealt{hunt14}, see also Sects. \ref{sec:atomicgas} -- \ref{sec:moleculargas})
the DGRs of these two galaxies 
differ by almost two orders of magnitude, despite their similar metallicity.

In this paper, we examine possible reasons for this dramatic difference in dust content
at similar metallicity in these two galaxies.
We first discuss their basic properties in Sect. \ref{sec:obs}.
Then, in Sect. \ref{sec:model},
we explore the origin of the observed dust in \sbs\ and \izw\ using a semi-analytic chemical
evolution model with dust. The model has been first introduced by \citet{valiante09} 
and then applied by \citet{valiante11,valiante12,valiante14} and \citet{debennassuti14} 
within the context of a hierarchical model for
the evolution of cosmic structures, as predicted by the concordance $\Lambda$CDM model. 
Here we do not investigate the hierarchical evolution of \sbs\ and \izw, but rather limit the discussion to their
chemical evolution assuming that they evolve in isolation with a constant star formation rate (SFR)\footnote{This assumption
is almost certainly inexact, but provides a benchmark for our assessment of consistency with observations.}. Below, 
we only briefly describe the model, referring interested readers to the above papers for a more detailed description.
We discuss implications of our results for high-redshift dust formation in star-forming galaxies in Sect. \ref{sec:discussion},
and our conclusions are given in Sect. \ref{sec:conclusions}.

\section{Observed properties of \izw\ and \sbs}
\label{sec:obs}

\begin{figure}
\vspace{\baselineskip}
 \includegraphics[width=\hsize]{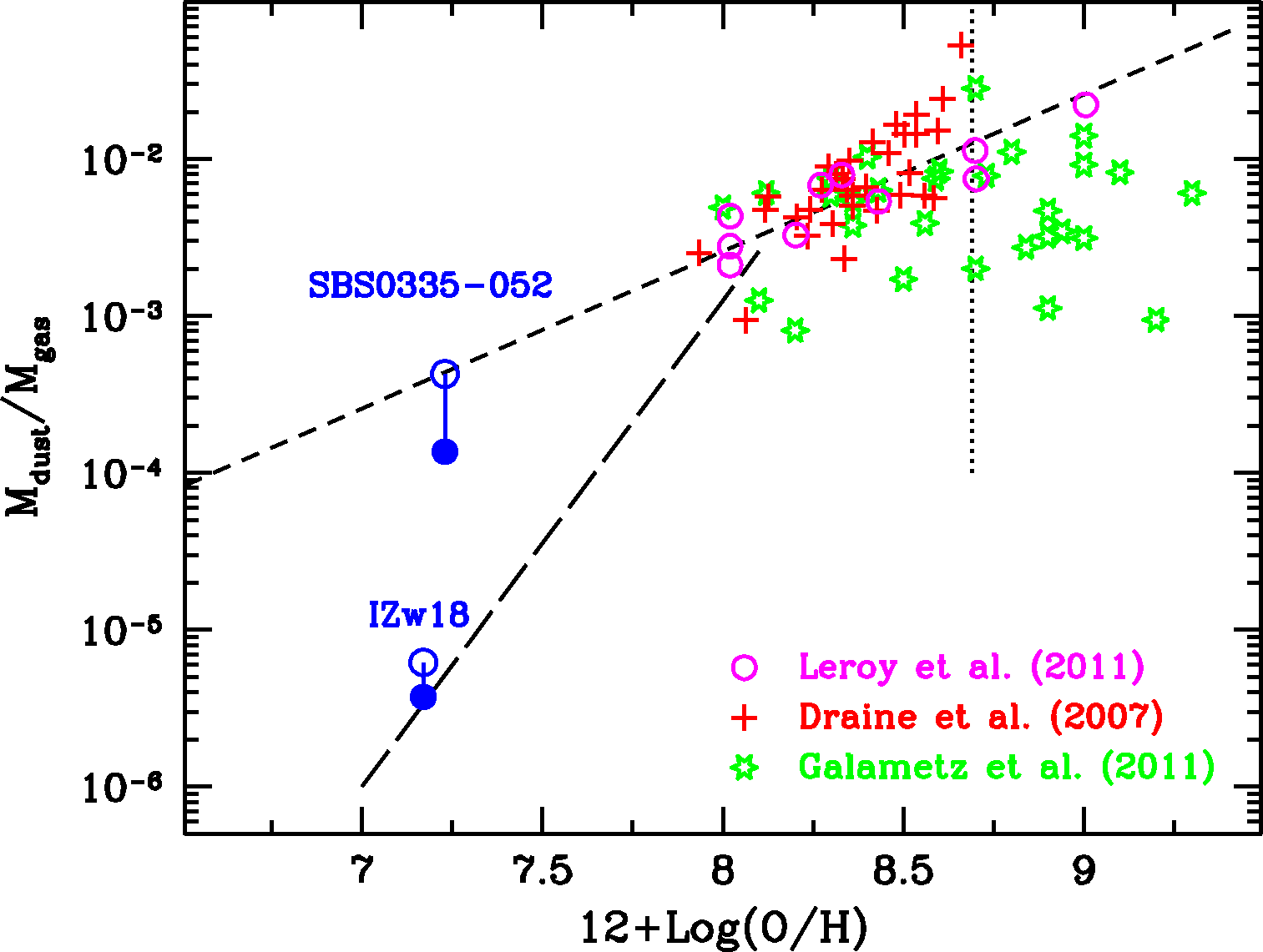}
 \caption{Dust-to-gas mass ratios (DGRs) as a function of metallicity.
Filled (blue) circles show DGRs inferred from the total gas mass of \sbs\ and \izw,
including ionized gas, atomic and (putative, as yet undetected) molecular gas;
open circles give DGRs including only the ionized and atomic gas components.
Galaxies from \citet{galametz11} are shown as (green) open stars,
those from \citet{draine07sings} as (red) crosses,
and from \citet{leroy11} as (magenta) open circles.
The short dashed line is the linear prediction by \citet{draine07sings}, and
the vertical dotted line illustrates solar metallicity \logoh\,=\,8.69 \citep{asplund09}.
The long dashed line (starting at \logoh\,=\,8.0) is the empirical fit
given by \citet{remy14} for low metallicities.
}
 \label{fig:dgroh}
 \end{figure}

The observed properties of the two metal-poor dwarf galaxies that we have adopted for our
analysis are presented in Table \ref{tab:data}. The observational data and the methods used to
infer the physical properties listed in the table have been thoroughly described in H14 
and references therein 
(see in particular Tables 1 and 3 in H14).
H14 analyzed dust and gas surface densities and did not include ionized gas.
Here we revise the estimates of gas masses in both galaxies relative to H14 in order to take into
account the ionized gas component and the spatial extent of cool dust emission.
The values obtained for the individual ISM components and the total gas masses are given
in Table \ref{tab:data} and shown graphically in Fig. \ref{fig:dgroh}.

\subsection{Atomic gas masses}
\label{sec:atomicgas}

Both \izw\ and \sbs\ are embedded in vast \hi\ envelopes which include other
galaxies or galaxy components. 
\sbs\ has a western component \sbs W at a distance
of $\sim$22\,kpc within a 64\,kpc \hi\ cloud \citep{pustilnik01,ekta09};
\izw\ (main body) lies about 2\,kpc away from the ``C'' component,
or Zwicky's flare,
within a diffuse \hi\ cloud extending over $\sim$19\,kpc \citep{vanzee98,lelli12}.
Thus it is difficult to determine exactly where the \hi\ of the galaxy
ends, and the more extended \hi\ envelope begins.
The large beams with which \hi\ is typically observed exacerbate the
problem, both because of beam dilution which causes \hi\ surface density to be
underestimated, but also because the tiny dimensions of the galaxies compared with
their \hi\ envelope make it difficult to assess the \hi\ content of the galaxy
itself.

Because our interest here is in dust production, and in the cool gas reservoir that provides the
fuel for star formation, we have taken \hi\ surface densities from the highest-resolution
observations available (as in H14), and calculated \hi\ mass over the optical extent of the
galaxy as determined from ($V$-band) surface-brightness profiles.
The $V$-band brightness profile of \izw\ falls to 25\,\magarc\ at a radius of
$\sim$8\,arcsec \citep[700\,pc,][]{hunt03}, and a radius $\sim$3\farcs2 for \sbs\ \citep[838\,pc,][]{thuan97ssc}.
With mean \hi\ surface densities \sigmahi\,=\,56\,\msunpc\ for \sbs\ \citep{thuan97} 
and 64\,\msunpc\ for \izw\ \citep{lelli12},
we thus estimate total \hi\ masses of 
3.1$\times 10^7$\,\msun\ and 2.5$\times 10^7$\,\msun\ 
for \sbs\ and \izw, respectively.
The \hi\ mass for \sbs\ is about 10 times lower than the total \hi\ mass by \citet{ekta09}
in a 40\,arcsec beam (about 6 times larger than the optical extent of the galaxy),
and $\sim$4 times lower than the total for \izw\ given by \citet{lelli12}.
Using the total \hi\ masses from the literature would lower the resulting DGR
by roughly these amounts \citep[e.g.,][]{remy14}.

\subsection{Ionized gas masses}
\label{sec:ionizedgas}

In low-metallicity star-forming galaxies such as \izw\ and \sbs, ionized
gas constitutes an important part of the total gas mass budget.
Hence, we have attempted to determine the mass of the ionized component in
the two galaxies.

Radio continuum observations of the free-free emission of ionized gas give
emission measures, and thus mean densities and source size.
The radio spectrum for \izw\ is flat, consistent with optically thin emission
\citep{cannon05,hunt05b}, while that of \sbs\ falls at low frequency,
indicative of free-free absorption and consequently high ionized-gas density \citep{hunt04,johnson09}.
As discussed in detail in Sect. \ref{sec:densities}, the density of
ionized gas in \izw\ is $\sim$5\,\cmthree, and $\sim$3200\,\cmthree\ in \sbs, corresponding
to emitting regions of $\sim$ 390\,pc radius for \izw\ \citep{hunt05b}
and $\sim$7.7\,pc radius for \sbs\ \citep{johnson09}.

Because ionized gas tends to be clumped 
\citep{kassim89,kennicutt84,zaritsky94,martin97,giammanco04,cormier12,lebout12}, we need to estimate a
volume filling factor over the optical extent of the galaxy.
We have done this by comparing the source size inferred from the radio emission measure 
to the optical size (see Sect. \ref{sec:atomicgas}), assuming a spherical geometry for the \hii\ region.
This comparison gives a volume filling factor of 0.17 for \izw\ 
and $\sim 10^{-6}$ for \sbs.
Although the value for \izw\ is within those observed for local \hii\ regions
\citep[e.g.,][]{kassim89}, the value for \sbs\ is extremely low.
This could be a consequence \citep[e.g.,][]{giammanco04} 
of the optically thick gas in this galaxy as indicated by the radio spectrum, 
so for \sbs\ we have used conservatively
a filling factor of $3\times10^{-4}$, roughly the lowest value found
by \citet{martin97}. 
With these densities and filling factors, and the optical size of the galaxy
as for \hi, we estimate ionized gas masses of
$3.1\times10^7$\,\msun\ and $5.9\times10^7$\,\msun\ for \izw\ and \sbs,
respectively.
The mass of the ionized gas is comparable to that of the atomic gas in both galaxies.

If, instead of the radio-determined value of 3200\,\cmthree, 
we consider the ionized gas density of 500\,\cmthree\ for \sbs\ inferred from the optical
[S {\sc ii}] lines \citep{izotov99}, and a larger filling factor of $10^{-3}$,
we would estimate an ionized gas density of $3\times10^7$\,\msun, 50\% of
our former estimate.
This can be considered as a rough measure of the uncertainty inherent in this calculation.

\subsection{Molecular gas masses}
\label{sec:moleculargas}

CO emission has never been detected in either galaxy;
thus it is difficult to measure \htwo\ masses, independently of the unknown
CO luminosity-to-\htwo\ mass conversion factor $\alpha_{\rm CO}$.
Thus, H14 measured the distance from the gas scaling relations in order to
estimate the missing (undetected) \htwo\ gas surface density
(for more details, see H14).
Using the densities from H14 (\sigmahtwo\,=94\,\msunpc\ and \sigmahtwo\,=\,342\,\msunpc,
for \izw\ and \sbs, respectively) and the optical size of the galaxy as above,
we infer \htwo\ masses for \izw\ and \sbs\ of
$3.6\times10^7$\,\msun\ and $1.9\times10^8$\,\msun, respectively.
These masses are highly uncertain, and assume that SFR surface density at low
metallicity follows the same scaling relations as more metal-enriched galaxies.

\subsection{Dust masses and comparison with previous work}
\label{sec:dustmass}

Dust masses for \sbs\ and \izw\ were measured by H14 by fitting the optical-to-mm
spectral energy distributions (SEDs) with spherical \dusty\ models \citep{ivezic97}.
Three different grain populations were included for determination of
the best-fit model (for more details see H14).
The resulting dust masses for both galaxies differ from those 
measured by other groups \citep[e.g.,][]{remy13,fisher14,remy14,izotov14}.
H14 discussed differences relative to \citet{remy13} for \sbs\ and \citet{fisher14} for \izw; 
here we compare our measurements to more recent work although
we are unable to compare our dust-mass estimates with \citet{remy14} because
there is no tabulation of their dust masses.

\citet{izotov14} use a multiple-temperature modified blackbody (MBB) approach, and 
fix the emissivity index for
all components. However, they only discuss the warm and cold dust (not
the hot) because, as they note, the hot dust (for $\lambda< 10\mu$m) is not in
thermal equilibrium. Considering their two-component fits, the SEDs shown in
their Fig. 6 almost never pass through the longer wavelength points. This is a
consequence of the assumption of single-temperature MBBs, and is expected to bias 
dust-mass estimates toward lower masses. Although seemingly a small
deviation, the constraints offered by longer wavelengths significantly raise
dust mass estimates, especially at low metallicity \citep[e.g.,][]{galametz11}.
It is well known that
including short wavelengths in a single-temperature MBB fit gives
unrealistically low dust masses because the mean temperature of the dust
radiating at these short wavelengths is higher than the bulk of the cooler
dust which dominates the mass. The warm dust dominates the light but the cool
dust dominates the mass and a single temperature 
\citep[or even two temperatures as in][]{izotov14} are not able to accommodate the temperature gradients. 

The cold-dust mass found by \citet{izotov14} for \izw\ is 70\%
of that found by H14;
compensating for the differences in dust emissivities  
worsens slightly the discrepancy to $\sim$65\%. 
This value is also a factor of 2 below the lower limit of the range of
dust mass estimated for \izw\ by \citet{fisher14} who use models by \citet{draine07},
even after compensating for the different emissivities adopted by the two groups.
More physically realistic models which
contemplate dust temperature gradients through variations in the
interstellar radiation field that heats the dust tend to give larger dust
masses than single-temperature MBB fits.

For \sbs, \citet{izotov14} find a cold dust mass of dex\,3.06\,\msun, roughly a factor
of 30 below H14. However, 
they estimate a ``cold" temperature of 57\,K, very close to the
temperature of 59\,K discussed by H14 under the assumption of single-temperature
dust, and thus do not take into account any cool dust. Moreover, the dust in
\sbs\ is optically thick at short wavelengths \citep{thuan99,plante02,houck04,hunt05a,hunt14};
thus the assumption of optically thin dust emission implicit in MBB fitting is incorrect.

The differences found in dust masses for \sbs\ and \izw\ of roughly a
factor of 100 are consistent with the differences in their integrated IR
luminosity: $L_{\rm IR}$ for \sbs\ is $1.6\times10^9$\,\lsun, while for \izw\
$L_{\rm IR}$\,=\,$2\times10^7$\,\lsun. 
Assuming the same mass-to-light ratio (similar overall mean temperatures) for
the dust, this would give a factor of 80 in dust mass, inconsistent with the
difference found by \citet{izotov14} of roughly a factor of 5 for the cold dust masses in
the two galaxies.  This illustrates one of the difficulties of the MBB
approach adopted by \citet{izotov14}; they use two temperatures and thus apply two
different mass-to-light ratios (roughly inverse temperature) to the two IR
integrals. As mentioned above, in galaxies like \sbs\ and \izw\ most of the dust
{\it emission} comes from warm dust, while most of the dust {\it mass} is cold.

More emissive grain mixtures, as suggested by \citet{jones13}, would decrease
the inferred dust mass by a factor 3-4 in both galaxies. 
A comparable reduction of the dust mass of \sbs\ 
($\sim 9 \times 10^3$\,\msun) would be obtained by artificially
imposing a low 870\,\micron\ dust flux (1$\sigma$ below the reported H14 flux).

We will discuss the implication
of this reduction in dust mass for our results in Sect.~\ref{sec:model}. 

Figure \ref{fig:dgroh} shows the DGRs of \izw\ and \sbs, together
with samples of galaxies taken from the literature (with dust masses obtained through
SED fitting rather than MBB approximations).
Because the dust-mass estimates for these samples are based on the \citet{draine07} models, 
the dust emissivities are comparable among the samples.
Figure \ref{fig:dgroh} also shows
the steeper-than-linear slope found by \citet{remy14} for galaxies with oxygen
abundance \logoh$\la$8.0.
The DGR of \sbs\ is roughly consistent with a linear slope of DGR with O/H \citep[e.g.,][]{draine07sings},
while \izw\ follows the steeper slope found by \citet{remy14}.

\begin{table*}
 \caption{Adopted physical properties of the two galaxies. The SFR values are
from radio-continuum SED fitting. The stellar ages are the mean values of stellar clusters and
the stellar masses have been derived integrating the SFR over the mean ages (see text).
The values in parentheses correspond to those from \dusty\ SED fitting (stellar mass, stellar age, metallicity, dust mass)
and to a constant SFR inferred from the \dusty\ SED best-fit stellar mass and ages. } 
 \label{tab:data}
\resizebox{\linewidth}{!}{
  \begin{tabular}{@{}lcrclllccccc@{}}
  \hline
\multicolumn{1}{c}{Name} & 
\multicolumn{1}{c}{Distance} & 
\multicolumn{1}{c}{Gas} & 
\multicolumn{1}{c}{$M_{\rm gas}$} & 
\multicolumn{1}{c}{$M_{\rm star}$} & 
\multicolumn{1}{c}{SFR} & 
\multicolumn{1}{c}{Age} & 
\multicolumn{1}{c}{12$+$} & 
\multicolumn{1}{c}{$Z$} & 
\multicolumn{1}{c}{$M_{\rm dust}$} &
\multicolumn{1}{c}{$n_{\rm mol}$} &
\multicolumn{1}{c}{$T_{\rm mol}$} \\
&  \multicolumn{1}{c}{(Mpc)}   & 
\multicolumn{1}{c} {Component} & 
\multicolumn{1}{c}{($10^7$\,\msun)}  & 
\multicolumn{1}{c}{$ (10^6$\,\msun)} & 
\multicolumn{1}{c}{(\msunyr)} & 
\multicolumn{1}{c}{(Myr)} & 
\multicolumn{1}{c}{Log(O/H)$^{\rm a}$} & 
\multicolumn{1}{c}{($\rm 10^{-2}$\,\zsun)} & 
\multicolumn{1}{c}{($10^2$\,\msun)} & 
\multicolumn{1}{c}{($\rm cm^{-3}$)} & 
\multicolumn{1}{c}{($\rm K$)} \\
\hline
\izw\         & 18.2 & { Ionized}   & {3.07}    \\ 
              &      & {Atomic}    & {2.46}    \\ 
              &      & {Molecular$^{b}$} & {3.62}    \\ 
              &      & {Total}     & {9.15}   & {1.02\,(1.82)} &  {0.17\,(0.10)} & {6\,(18.3)} & {7.18} & { 3.09\,(2)} & $(3.4 \pm 1)$  &  100  & 10 \\
\sbs\         & 54.1 & {Ionized}   & {5.85}    \\ 
              &      & {Atomic}    & { 3.09}    \\ 
              &      & {Molecular$^{b}$} & {18.9}    \\ 
              &      & {Total}     & {27.8}  & {7.92\,(23.5)}  &  {1.32\,(1.79)} & {6\,(13.1)} & {7.27} & { 3.80\,(2)} & $(3.8 \pm 0.6)\times 10^2$ & 1500 & 80 \\
\hline
\end{tabular}
}
\flushleft{ $^{\rm a}$~Averages taken from \citet{izotov99}. \\
$^{\rm b}$~\htwo\ surface densities inferred by H14 from gas scaling relations, not directly detected. See text for more details.
}
\end{table*}

\subsection{Stellar masses, stellar ages, and SFR }
\label{sec:mass}

Stellar masses in star-forming dwarf galaxies such as \sbs\ and \izw\ are notoriously
difficult to determine.
The main problem is contamination by nebular continuum emission which affects both
the optical \citep[e.g.,][]{reines08,reines10,adamo10} and the
near-infrared emission \citep[e.g.,][]{smith09,hunt12}.
In \sbs, the contamination from free-free emission at 3.4\,\micron\ is 27\% \citep{hunt01}
and $\sim$50\% at 2.2\,\micron\ \citep{vanzi00};
even more extreme contamination is observed in \izw, with $>$50\% of the IRAC 4.5\,\micron\
flux due to nebular emission \citep{hunt12}.
Hot dust is also a problem, especially in \sbs\ where it comprises $\sim$67\% of its 4\,\micron\ emission \citep{hunt01}.
While nebular continuum levels can be estimated from SFRs
\citep[e.g.,][]{smith09,hunt12}, it is difficult to ascertain hot-dust levels without detailed
multi-wavelength photometry.
Thus stellar masses of low-metallicity dwarf starbursts such as \sbs\ and \izw\ are
prone to large uncertainties.

Table \ref{tab:data} reports in parentheses the values of stellar mass, \mstar, and age inferred from
\dusty\ SED fitting (H14), and SFRs that correspond to the constant values needed to produce the
\dusty\ SED best-fit \mstar when integrated over the best-fit age. 
However, for our analysis, we require a {\it mean} age for the stellar populations producing the dust currently observed.
A mean age of stellar clusters of 6 Myr has been calculated by averaging values from \citet{recchi02} and \citet{hunt03} for \izw\ and from
\citet{reines08} for \sbs. Thus, we have also computed the stellar mass  accumulated over the timespan
of the mean ages of the clusters at a constant rate given by the observed values of SFR
derived from radio free-free emission \citep[][]{hunt05b,johnson09}. All these values (not in parentheses) are also reported in Table \ref{tab:data}.

The resulting \mstar\ values are within 40\% of 
previous estimates obtained by fitting the optical-NIR SEDs of individual
star clusters with single stellar population models
\citep[\sbs: $5.6\times10^6$\,\msun, \izw: $7.0\times10^5$\,\msun,][respectively]{reines08,fumagalli10}.

Only if age priors (3-6\,Myr) are imposed for the fit are
the \dusty\ stellar masses obtained by H14 
consistent with these values for \sbs.
There is a similar discrepancy for the \dusty\ stellar mass of \izw\ at an age of 18.3\,Myr,
almost a factor 2 larger than the value obtained with a constant SFR of
0.17\,\msunyr. 
Part of the reason for this is that, at a given luminosity, the mass-to-light ratios are smaller for younger
stellar populations (see H14 for details).
Another reason is that there is an age gradient in both galaxies, and the
SED fitting relies on global photometry that encompasses both young and old clusters. 
In \sbs, the northern super-star clusters (SSCs) are older with a maximum age of $\sim$12\,Myr, compared to $\la$3\,Myr
for the southern ones \citep{reines08,adamo10};
in \izw, the SE cluster and C component are older ($\sim10-15$\,Myr) compared to the younger NW
cluster \citep[$\sim$3\,Myr,][]{hunt03}.

\begin{figure}
 \includegraphics[width=\hsize]{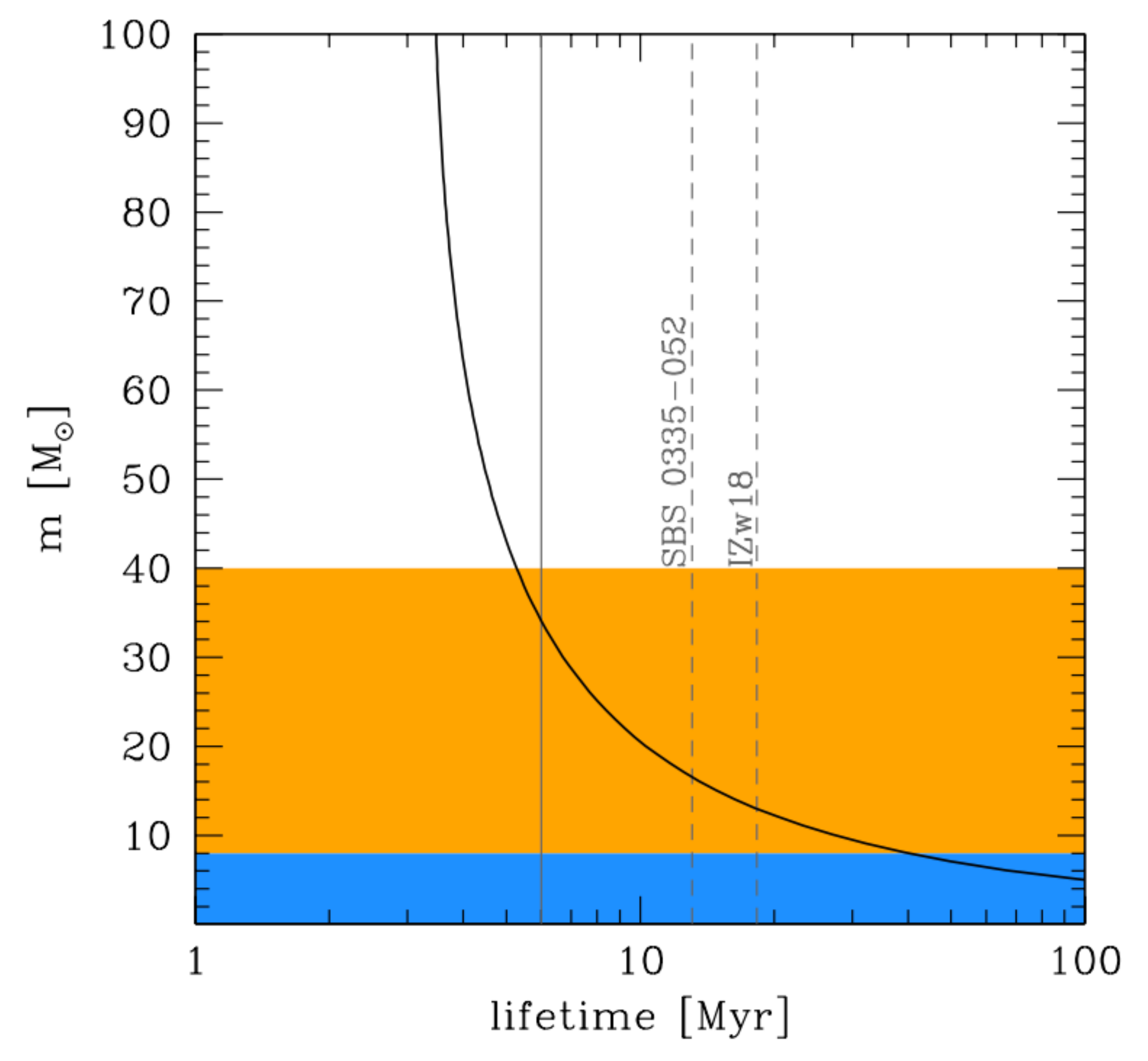}
 \caption{The dependence of the stellar lifetime on the initial stellar mass for stars with metallicity \zzsun\,=\,0.03 
(solid black line). 
 The orange and  blue regions indicate the progenitor mass ranges of core-collapse SN and AGB stars, respectively. 
 The solid and dashed
 vertical lines indicate the estimated stellar ages for the two galaxies, based on individual super stellar clusters fits and on \dusty\ SED fits, respectively (see text).}
 \label{fig:stellartime}
 \end{figure}

Such a difference in the stellar ages can have important implications for the
chemical evolution of the system. In Fig.~\ref{fig:stellartime} we show the stellar mass-lifetime
relation for stars with initial metallicity \zzsun\,=\,0.03, 
using the \citet{raiteri96} 
formulation. The shaded regions illustrate the ranges of stellar masses that contribute to metal
enrichment by means of core-collapse SN explosions and stellar winds from intermediate
mass stars. The vertical grey lines indicate the values of 
stellar ages shown in the table (dashed lines are the \dusty\ stellar ages). 
The figure shows that part of the metals and dust that we 
presently observe may have originated {\it in situ}, from supernova explosions ({\it self-enrichment} scenario).
However, if we adopt the mean age of the stellar clusters (solid line), only the most
massive supernovae with 34\,\msun\,$\le\,m\,\le$\,40\,\msun \, have evolved to their metal production stage. 
Hence the ISM of the galaxies must have achieved most of its metal content prior to the 
current star-formation episode
({\it pre-enrichment} scenario). 
 If, instead, the stellar population age for \sbs \, (\izw) is 13.1\,Myr (18.3\,Myr), as inferred from \dusty\ SED fit,
then the mass range of the stars that can contribute to the ISM enrichment extends to 16\,\msun\,$\le\,m\,\le$\,40\,\msun\,
(13\,\msun\,$\le\,m\,\le$\,40\,\msun). 

\begin{figure}
\includegraphics[width=\hsize]{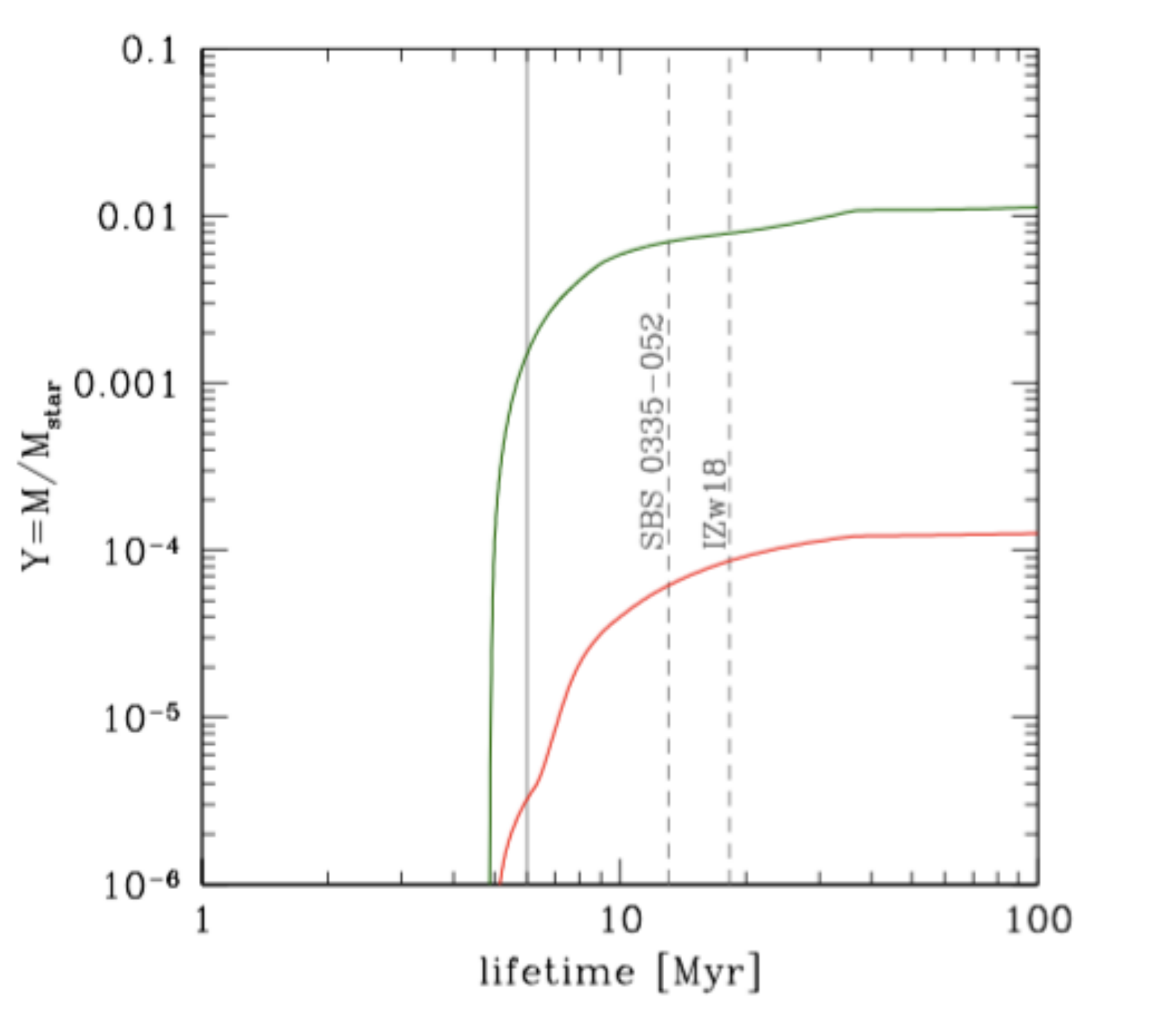}
 \caption{The mass fraction of metals (dark green) and dust (red) relative to the mass of stars as a function of the stellar age. 
 The solid and dashed
 vertical lines indicate the estimated stellar ages for the two galaxies, based on individual super stellar clusters fits and on \dusty\ SED fits, respectively (see text). }
 \label{fig:cosmicyields}
 \end{figure}

Fig.~\ref{fig:cosmicyields} quantifies this difference in terms
of the mass of metals and dust that can be produced by self-enrichment. It shows that -- depending
on the adopted stellar age -- the mass fraction of metals (dust) relative to the stellar mass is in the range 
$1.5\times 10^{-3} \le Y_{\rm Z} \le 7 \times 10^{-3}$  
($3.26\times 10^{-6} \le Y_{\rm d}  \le 6.2 \times 10^{-5}$)
for \sbs \, and in the range $1.5\times 10^{-3} \le Y_{\rm Z} \le 7.9 \times 10^{-3}$  
($3.26\times 10^{-6} \le Y_{\rm d} \le 8.6 \times 10^{-5}$) for \izw. Using the
corresponding stellar masses reported in Table~\ref{tab:data}, we find that -- even assuming that all the newly
formed dust injected in the ISM is conserved -- the dust mass  produced by self-enrichment is always smaller than
observed, being $26 \, M_\odot \le M_{\rm d} \le 1460 \, M_\odot$ for
\sbs \, and $3 \, M_\odot \le M_{\rm d} \le 157 \, M_\odot$ for \izw.
As explained below, these values have been obtained
using the dust and metal yields presented by \citet{valiante09} 
and assuming that the stars form at a constant rate with a Salpeter Initial Mass Function (IMF) in a mass range
0.1\,\msun\,$\le\,m\le$ 100 \,\msun\ 
with a metallicity of $Z = 0.03 \, Z_\odot$.

\subsection{Cold gas densities and temperatures}
\label{sec:densities}

The remaining quantities required for our models
are the mean density and temperature of the cold neutral medium (CNM) and molecular
clouds in which dust grains form and grow by accretion (see Sect.\,\ref{sec:model}).
We can derive the number densities of the molecular phase by assuming thermal pressure balance at
the atomic-molecular interface \citep[e.g.,][]{krumholz09a}.
Observationally and from a theoretical point of view,
the mean number density of molecular clouds,
\meannhtwo, is expected to be $ \phi_{\rm mol} \sim 10$ times that in the CNM of the ISM: 
\meannhtwo\,=\,$\phi_{\rm mol} \, \langle n_{\rm CNM} \rangle$.
We can therefore calculate approximate values of $n_{\rm mol}$ in \sbs\ and \izw\
by first considering the observed \hi\ column densities\footnote{We have considered regions of
similar size, constrained by requiring a similar physical resolution of the \hi\ observations: $\sim$524\,pc (2\arcsec) for \sbs\
and $\sim$436\,pc (5\arcsec) for \izw.}:
$7\times10^{21}$\cmtwo\ and
$4.8\times10^{21}$\cmtwo\ for \sbs\ and \izw, respectively \citep{thuan97,lelli12}. 
These can be converted to mean volume densities over the regions of interest
(the area subtended by the massive star clusters)
by considering the diameters of the star-forming region: $\sim \,$15.8\,pc for \sbs\ \citep{johnson09}
and $\sim \,$170\,pc for \izw\ \citep{cannon02,hunt09}. 
We thus obtain a mean molecular density \meannhtwo$\, \sim \,$1435\,\cmthree \, and \meannhtwo$\, \sim \,$91\,\cmthree\ for \sbs\ and \izw, respectively.

Despite the similar metallicities of the two galaxies, the inferred molecular densities differ 
by more than an order of magnitude.
Nevertheless, these estimates are 
roughly consistent, given the uncertainties, with the 
densities that would be inferred from the putative (unobserved) molecular component discussed by H14. 
Assuming that the Kennicutt-Schmidt law relating gas and star-formation surface densities holds also
for these two galaxies, and with the same sizes as above, we would derive
\meannhtwo$\,\sim \,$1800\,\cmthree\ for \sbs\ and \meannhtwo$\,\sim \,$40\,\cmthree\ for \izw.
The radio spectrum also shows evidence for such a difference in gas densities in the two galaxies.
While in \sbs\ there is the clear signature of a strongly absorbed thermal component, both
globally \citep{hunt04} and around the individual southern SSCs \citep{johnson09}, 
\izw\ shows a typical flat brem\ss trahlung spectrum \citep{hunt05b}.
Based on fits of the radio spectrum,
the ionized-gas densities inferred for the two objects are $\sim$3200\,\cmthree\ for \sbs\ \citep{johnson09},
and $<$10\,\cmthree\ for \izw\ \citep{hunt05b}.
Finally, even the optical spectra of the two objects show a large difference in the densities
measured from the [S{\sc ii}] lines: $\ga$500\,\cmthree\ for \sbs\
and $\la$100\,\cmthree\ for \izw\ \citep{izotov99}.
Given the uncertainties in the above estimates, in what follows we take the reference values of 1500\,\cmthree\
and 100\,\cmthree\ for the molecular gas densities in \sbs\ and \izw, respectively.

Following \citet{krumholz09a}, we can use the temperature-density relation for the CNM predicted by \citet{wolfire03}
ISM model\footnote{We have considered
the most general expression for the CNM temperature, which takes into account
the vastly different DGRs for these two galaxies.} 
to find $T_{\rm CNM}$ and then derive the temperature of the molecular gas implied by thermal pressure balance,
$T_{\rm mol} = 1.8 \, T_{\rm CNM}/\phi_{\rm mol}$.
The resulting values for the two galaxies are shown in Table \ref{tab:data}.

\section{The chemical evolution model}
\label{sec:model}

The equations describing the chemical evolution of a galaxy that evolves in isolation (closed-box approximation),
can be summarized as follows:

\begin{equation}
\label{eq:starevo}
\dot{M}_{\rm star}(t) =  {\rm SFR}(t) - \dot{R}(t),
\end{equation}
\begin{equation}
\label{eq:ISMevo}
\dot{M}_{\rm ISM}(t) =  - {\rm SFR}(t) + \dot{R}(t)
\end{equation}
\begin{equation}
\label{eq:metevo}
\dot{M}_{\rm Z}(t)  =  - Z_{\rm ISM}(t) {\rm SFR}(t) + \dot{Y}_{\rm Z}(t)
\end{equation}
\begin{eqnarray}
\label{eq:dustevo}   
 \dot{M}_{\rm d}(t)  & = &  - Z_{\rm d}(t) {\rm SFR}(t) + \dot{Y}_{\rm d}(t) \\ \nonumber
 & & - (1- X_{\rm c}) \frac{M_{\rm d}(t)}{\tau_{\rm d}}+ X_{\rm c} \frac{M_{\rm d}(t)}{\tau_{\rm acc}}
\end{eqnarray}
\noindent
where $M_{\rm star}$ is the stellar mass, $M_{\rm ISM}$ is the total mass in the ISM (the sum of the gas and dust masses), $M_{\rm Z}$ is the 
total mass in heavy elements (both in the gas phase and in dust grains), $M_{\rm d}$ is the dust mass (so that the mass of gas phase
elements is given by $M_{\rm met} = M_{\rm Z} - M_{\rm d}$), $Z_{\rm ISM}(t)=M_{\rm Z}(t)/M_{\rm ISM}(t)$ is the ISM metallicity, and
$Z_{\rm d}(t)=M_{\rm d}(t)/M_{\rm ISM}(t)$ is the total dust abundance in the ISM. The terms $\dot{R}(t)$, $\dot{Y}_{\rm Z}(t)$ and $\dot{Y}_{\rm d}(t)$  
are the rates at which the mass of gas, heavy elements and dust is returned to 
the ISM after stellar evolution, respectively. These time-dependent terms
depend on the adopted model grids and stellar IMF. We compute them as follows:
\begin{equation}
\dot{R}(t) =  \int_{m(t)}^{m_{\rm up}} (m - m_{\rm r}(m,Z))\ \Phi(m)\ {\rm SFR}(t - \tau_{\rm m})\ dm,
\label{eq:returnedgasmass}
\end{equation}
\begin{equation}
\dot{Y}_{\rm Z}(t) = \int_{m(t)}^{m_{\rm up}} m_{\rm Z}(m,Z)\ \Phi(m)\ {\rm SFR}(t - \tau_{\rm m})\ dm,
\label{eq:metalyields}
\end{equation}
\begin{equation}
 \dot{Y}_{\rm d}(t) = \int_{m(t)}^{m_{\rm up}} m_{\rm d}(m,Z)\ \Phi(m)\ {\rm SFR}(t - \tau_{\rm m})\ dm,
\label{eq:dustyields}
\end{equation}
\noindent
where the lower limit of integration, $m(t)$, is the mass of a star with a lifetime $\tau_{\rm m} = t$; $m_{\rm r}, \, m_{\rm Z},$ and $m_{\rm d}$ are  
respectively the remnant mass, the metal and dust mass yields, which depend on the stellar mass and 
metallicity; and $\Phi(m)$ is the stellar IMF, which we assume to be a Salpeter law in the mass range 0.1\,\msun$\le m \le$\,100\,\msun\ \citep[e.g.,][]{valiante09}.
For stars with $m\,< 8 $\,\msun, 
we adopt metal yields from
\citet{vandenhoek97} 
and dust yields for intermediate-mass stars on the AGB phase of the evolution by 
\citet{zhukovska08}.  
For massive stars (12\,\msun\,$<\,m\,<$\,40\,\msun) 
metal and dust yields have been taken from 
\citet{woosley95} 
and from \citet{bianchi07}, 
using the proper mass-and metallicity-dependent values and including the effect of the reverse shock on dust survival
in SN ejecta. 
For stars in the intermediate mass range 8\,\msun\,$\le \,m\,\le $\,12\, \msun, 
we interpolate between the AGB yields 
for the largest-mass progenitor and SN yields from the lowest-mass progenitor. 
Above 40 \msun, stars are assumed to collapse to black hole without contributing to the enrichment of the ISM. 
Finally, the last two terms in the right-hand side of eq.~(\ref{eq:dustevo}) represent the effects of dust destruction by interstellar
shock waves and grain growth in the dense phase of the ISM. Here we simplify the treatment of the two-phase ISM model
described in \citet{debennassuti14}, 
and quantify the fraction of ISM in the cold phase with the parameter $X_{\rm c}$.
This is taken to be time-dependent and rescaled from the SFR,
\begin{equation}
{\rm SFR}(t) = \frac{\epsilon_\ast X_{\rm c} (t) M_{\rm gas}(t)}{\tau_{\rm ff}},
\label{eq:coldphase}
\end{equation}
\noindent
where $\epsilon_\ast$ is the star formation efficiency \citep{krumholz12}, 

\begin{equation}
\tau_{\rm ff} = \sqrt{\frac{3 \pi}{64\, G\, m_{\rm H} \, <n_{\rm mol}>}} 
\end{equation}
\noindent
is the free-fall timescale at the mean density of molecular clouds $<\rho_{\rm mol} > \sim 2 \, m_{\rm H} \, <n_{\rm mol}>$.
Finally, the timescale for grain destruction, $\rm \tau_{d}$ and grain growth $\rm \tau_{acc}$ are computed as in 
\citet{debennassuti14}. 
The latter timescale
depends on the gas phase metallicity, temperature and density of molecular clouds,

\begin{equation}
\tau_{\rm acc} = 20 \, {\rm Myr} \times \Big( \frac{n_{\rm mol}}{100\, {\rm cm}^{-3}} \Big)^{-1}\, \Big( \frac{T_{\rm mol}}{50\, {\rm K}} \Big)^{-1/2}\, \Big( \frac{Z}{Z_\odot} \Big)^{-1}
\end{equation}
\noindent
where we have assumed that grains which experience grain growth have a typical  size of $ \sim 0.1\,$\micron\ \citep{hirashitavosh14}. 
If $n_{\rm mol} = \rm 10^3 cm^{-3}$ and $T_{\rm mol} = \rm 50 \, K$, the accretion timescale for gas at solar metallicity is only 2 \,Myr \citep{asano13}. 

Below we apply the chemical evolution model to each of the two galaxies under investigation.  Throughout the following, we adopt a solar metallicity of $Z_\odot = 0.0134$
\citep{asplund09}.

\subsection{\sbs}

The observed mass of metals  in \sbs \, is $M_{\rm met}(t_{\rm obs}) = Z  \, M_{\rm gas}(t_{\rm obs})  =   [4.53\times 10^4 \,  - \, 1.41 \times 10^5] \, \, M_\odot$
(the lower limit corresponds to pure atomic gas and the upper one to the total gas mass, see Table~\ref{tab:data}). The observed dust mass is
$M_{\rm dust}(t_{\rm obs}) = 3.8 \times 10^4 \, M_\odot$.  
Hence, the total mass in heavy elements $M_{\rm Z}(t_{\rm obs})$ ranges from $8.33 \times 10^4 \, M_\odot$ to $1.79 \times 10^5 \, M_\odot$. 
As described in Sect.~\ref{sec:obs}, the maximum dust mass that can be produced by self-enrichment from observed stellar populations in \sbs\, is
$\sim 1.5 \times 10^3 \, M_\odot$, $\sim 25$ times lower than the estimated dust mass reported in Table~\ref{tab:data} \citep{hunt14} and $\sim 50\%$ lower
than the lowest limit on the dust mass reported by \citet{thuan99}. It is therefore unlikely that the origin of the dust in \sbs\ is self-enrichment. 
In fact, the same stars 
responsible for metal pre-enrichment may have also formed dust. Alternatively, the dust mass may have formed
{\it in situ} by means of grain growth. We examine each of these two possibilities in turn.

We first assume that the stars observed in \sbs, with a total mass of 
$M_{\rm star} =  7.92 \times 10^6\,  M_\odot$, have a mean age of 6 Myr and have formed at a constant rate of 
$1.32\, M_\odot$/yr with a Salpeter IMF (see the values in Table \ref{tab:data}). 
Under these conditions, only stars with masses 
$m \ge 34 \, M_\odot$ had the time to evolve and the total mass of heavy elements and dust injected in the ISM by SNe is  $\rm 1.18 \times 10^4 \, M_\odot$
and $\rm 25.8 \, M_\odot$, respectively, much smaller than observed.
From eq.~(\ref{eq:metevo}), we can estimate the mass of metals that was originally present in \sbs,  $M_{\rm Z}(t_{\rm ini})$, where $t_{\rm ini} = t_{\rm obs} - 6\, {\rm Myr}$.
This is equal to the mass presently observed, corrected for astration and self-enrichment, 
\begin{equation}
M_{\rm Z}(t_{\rm ini}) = M_{\rm Z} (t_{\rm obs}) - \int_{t_{\rm ini}}^{t_{\rm obs}} dt \, \left[-Z_{\rm ISM}(t) {\rm SFR}(t) + \dot{Y}_{\rm Z}(t)\right],
\label{eq:premetSBS}
\end{equation}
\noindent
and we find $M_{\rm Z}(t_{\rm ini})\sim  {1.79 }\times 10^5 \, M_\odot$.
Assuming the Salpeter IMF-averaged dust and metal yields for a fully-evolved stellar population with a metallicity 
$Z = 0.03 \, Z_\odot$ ($Y_{\rm Z} = 1.43\times 10^{-2}$ and $Y_{\rm d} = 4.6 \times 10^{-4}$), the
maximum mass of dust that \sbs\ could have achieved by pre-enrichment can be estimated as
\begin{equation}
M_{\rm d}(t_{\rm ini}) = \frac{Y_{\rm d}}{Y_{\rm Z}} \, M_{\rm Z}(t_{\rm ini}) \sim  { 5.7 \times 10^3} M_\odot,
\label{eq:predustSBS} 
 \end{equation} 
\noindent
which is only  $\sim 15\%$ of the observed value. 

We can repeat the same calculation but assuming that the observed stars in \sbs \, have a mean age of 13.1~Myr,  a total
stellar mass of $M_{\rm star} =  2.35 \times 10^7 \, M_\odot$ and have formed with a constant SFR of $\rm 1.79 \, M_\odot$/yr with a 
Salpeter IMF (see the values in parenthesis in~Table~\ref{tab:data}). In this case, newly formed stars 
with masses $m \ge 16 \, M_\odot$ produce $1.65 \times 10^5 \, M_\odot$ of heavy elements and $1460 \, M_\odot$ of dust, too
small to account for the observed dust  mass.  Using eq.~(\ref{eq:premetSBS}), 
we find that the mass of heavy elements achieved by means of pre-enrichment at $t_{\rm ini} = t_{\rm obs} - 13.1 \, {\rm Myr}$
is $M_{\rm Z}(t_{\rm ini}) \sim 1.56 \times 10^5 \, M_\odot$, which corresponds to a maximum dust mass of $M_{\rm d}(t_{\rm ini}) \sim 5 \times 10^3 \, M_\odot$ 
(from eq.~\ref{eq:predustSBS}), $\sim 13 \%$ of the observed value. 

It is important to stress that the above values of $M_{\rm d}(t_{\rm ini})$ should be regarded as upper mass limits.
In fact, if we were to consider only the atomic gas rather than the total (including the putative molecular component), the dust masses
allowed by pre-enrichment would be even lower, $\lesssim 10\%$ in both cases.  Moreover,
eq.~(\ref{eq:predustSBS}) is based on the implicit assumption that all the dust produced by
previous stellar generations has been conserved in the ISM, without undergoing any destruction by interstellar shocks.
Hence, pre-enrichment cannot account for the observed dust masses, 
even if we were to consider a factor 3-4 reduction in the observed dust mass,
either by using more emissive grain mixtures \citep{jones13},
or by artificially lowering the observed 870\,\micron\ dust flux in the SED
(see Sect. \ref{sec:dustmass}).

Independently of the adopted stellar ages, a major fraction of the existing mass of dust in \sbs\ must have formed by means of
grain growth in the dense phase of the ISM.  

In Figure \ref{fig:SBSS0335evo} we show the results of chemical evolution models. 
Since it is impossible to constrain the initial value of dust mass that \sbs\ has inherited from previous stellar generations,
we have explored two limiting cases: in the first one, we take the observed gas density of  \meannhtwo$\sim$1500\,\cmthree\ 
(see Table \ref{tab:data}) and we start from the minimum initial dust mass that allows the model to reproduce the observations (red shaded region between the
two solid lines).
This model shows that if the stellar age is 6~Myr, \sbs\ must start with the maximum possible initial dust mass predicted 
by eq.~(\ref{eq:predustSBS}), grain growth accounts for $\sim 85 \%$ of the existing dust mass, with pre-enrichment
providing the remaining $\sim 15\%$. If the stellar age is 13.1\,Myr, due to the longer time available for grain growth,
\sbs \, can start with a dust mass that is $\sim 30\%$ of the maximum value achieved by pre-enrichment and grain growth
accounts for $\sim 95 \%$ of the observed dust mass. 
In the second model (blue shaded region between the two shaded lines), we assume an initial dust mass of only  $1 \, M_\odot$, as if the chemical initial 
conditions inherited from previous stellar generations were not favourable to dust pre-enrichment. Under this pessimistic scenario,
a gas density of $n_{\rm mol} = 10^4 \rm \, cm^{-3}$ ($5 \times 10^3 \rm \, cm^{-3}$) would allow grain growth to increase the dust mass, reaching the observed value in 6\,Myr (13.1\,Myr). 
The above results have been obtained solving the system of equations (1) - (4) assuming an initial gas mass $M_{\rm gas}(t_{\rm ini}) = M_{\rm gas}(t_{\rm obs}) +  M_{\rm star}(t_{\rm obs})$
(closed-box approximation) and fixing the free parameters $X_{\rm c}$ and $\epsilon_\ast$ to reproduce the mass of molecular gas component at $t_{\rm obs}$. For 
\meannhtwo$\sim$1500\,\cmthree\, this constrains the star formation efficiency to be $7 \times 10^{-3} \leq \epsilon_\ast \leq 9 \times 10^{-3}$ and the fraction of molecular
gas to be $X_{\rm c} = 0.67$. 

Hence we conclude that, provided that a major fraction of the dense gas in \sbs\ is at densities 
$1.5 \times 10^3 \,$cm$^{-3} < n_{\rm mol} < 10^4\,$cm$^{-3}$, the observed dust mass in \sbs \, can be 
reproduced by means of grain growth. \\

\subsection{\izw}
We next analyse the evolution of \izw. The observed mass of metals and dust are  
$M_{\rm met}(t_{\rm obs}) = Z \, M_{\rm gas}(t_{\rm obs}) = [2.29 - 3.78] \times 10^4 \, M_\odot$ (the lower limit
corresponds to pure atomic gas and the upper one to the total gas mass, see Table \ref{tab:data}) and 
$M_{\rm d}(t_{\rm obs}) = 3.4 \times 10^2 \, M_\odot$, resulting in a total mass of heavy elements of $M_{\rm Z}(t_{\rm obs}) = [2.32 - 3.82] \times 10^4 \, M_\odot$. 
Depending on the adopted stellar ages ($\rm 6\, Myr \leq t_{age} \leq 18.3 \, Myr$), between 4 and 60\% of the observed metal mass can
be achieved by self-enrichment, the remaining fraction must have come from previous stellar generations. The same is true for the dust mass:
even assuming that all the dust injected by SNe in the ISM is conserved, self-enrichment can produce between $\lesssim 1$ and $\sim 46\%$
of the existing dust mass. 
Using eqs.~(\ref{eq:premetSBS}) and (\ref{eq:predustSBS}), the initial mass of heavy elements and dust can be estimated to be:
\begin{eqnarray}
M_{\rm Z}(t_{\rm ini})  & \sim & (1.93 - 3.82) \times 10^4 M_\odot, \nonumber \\
M_{\rm d}(t_{\rm ini}) & = &  \frac{Y_{\rm d}}{Y_{\rm Z}} \, M_{\rm Z}(t_{\rm ini}) \sim   (0.62 - 1.23) \times 10^3 M_\odot, \nonumber
 \end{eqnarray} 
\noindent 
where the upper (lower) initial values have been obtained assuming a stellar age of 6 (18.3)\,Myr and considering only
atomic versus total gas mass.
Because the initial maximum dust masses are a factor 2 -- 4 larger than the observed value, all of the existing dust mass in \izw\,
could originate from dust pre-enrichment. 

\begin{figure}
\includegraphics[width=\hsize]{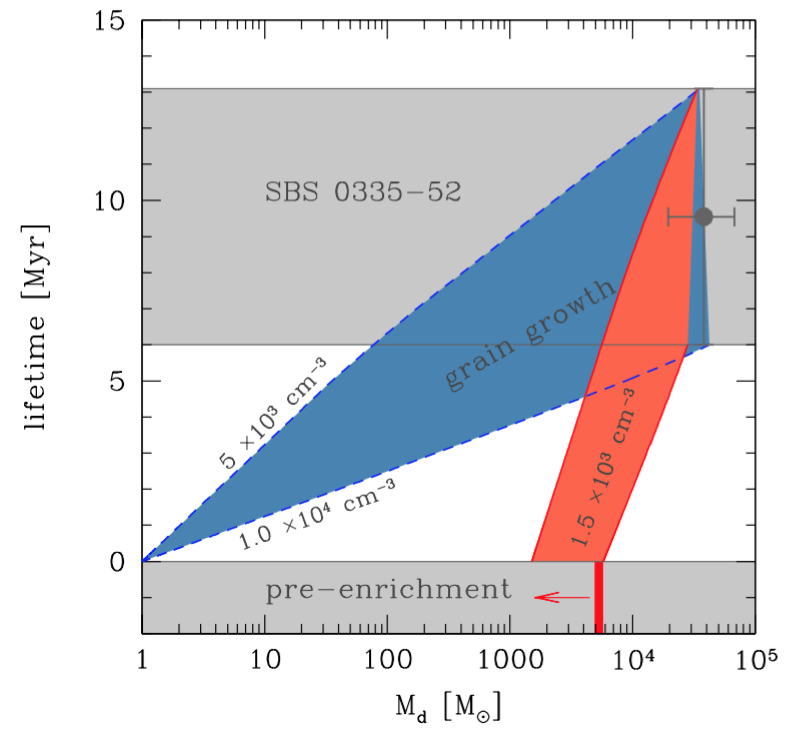}
 \caption{The predicted evolution of the dust mass in \sbs \, is compared with the observed value (grey data point). 
 The upper grey shaded region indicate the range of variation of the inferred stellar ages.
 The maximum initial dust mass achieved by pre-enrichment is illustrated by the solid red line in the lower grey
 shaded region. The red region illustrate the range of models where the observed dust mass is
 reproduced adopting  \meannhtwo$\sim$1500\,\cmthree \, to compute the grain growth timescale 
 and requiring an initial dust mass from pre-enrichment. The blue region
 illustrate models where dust pre-enrichment is negligible and 
 we adjust the density of dense gas so that the observed dust mass is reproduced by means of grain growth (the
 required range of $\rm n_{mol}$ is also shown, see text).}
 \label{fig:SBSS0335evo}
 \end{figure}

\begin{figure}
\includegraphics[width=\hsize]{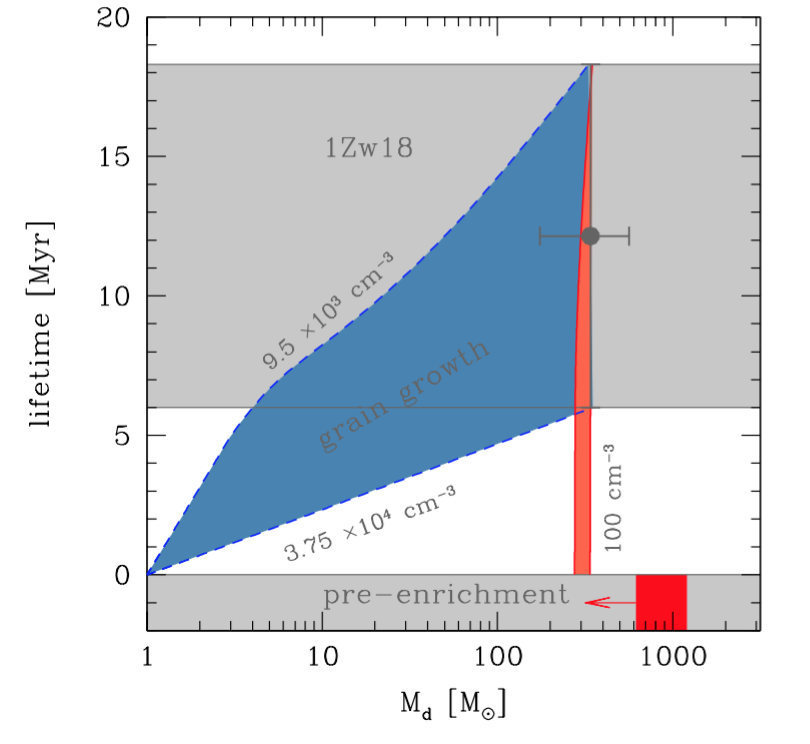}
 \caption{Same as Fig.\ref{fig:SBSS0335evo} but for \izw.}
 \label{fig:IZw18evo}
 \end{figure}

Figure \ref{fig:IZw18evo} shows the results of chemical evolution models. 
Similarly to the case of \sbs, we solve the system of equations (1) - (4) in the closed-box approximation with 
parameters $X_{\rm c}$ and $\epsilon_\ast$ fixed to reproduce the mass of molecular gas component at $t_{\rm obs}$. 
We first fix the gas density to
its observed value, \meannhtwo\,=\,100\,\cmthree, and we run the model starting from the minimum initial dust mass that allows the
observed dust mass to be reproduced (red shaded region between the two solid lines). This implies a star formation efficiency 
$1.7 \times 10^{-2} \leq \epsilon_\ast \leq 10^{-2}$ and a molecular gas fraction of $X_{\rm c} = 0.39$. 
It is clear that grain growth is negligible and 
that the required initial dust mass  is smaller when the stellar age is 18.3\,Myr, 
due to the larger contribution given by self-enrichment. 
According to this scenario, the system starts with an initial dust mass of $M_{\rm d}(t_{\rm ini}) \sim [276 - 338] \, M_\odot$, and
then evolves very slowly, with dust injected by newly-formed stars, partly compensated by grain destruction by interstellar shocks
and astration. As a result, more than $\sim 80\%$ 
of the observed dust mass is inherited from pre-enrichment, with grain growth making up the rest. This is different from the solid lines
shown in Fig.~\ref{fig:SBSS0335evo}, and a consequence of the lower gas density observed in \izw, which causes grain growth
to be inefficient.  
The dashed lines in Fig.~\ref{fig:IZw18evo} show the evolution when -- by construction -- we impose grain growth to provide
the dominant contribution to the final dust mass. In this case, we can start from a negligible initial dust mass but 
the required gas densities are in the range 
 $9.5 \times 10^3$\cmthree\,$\le \rm n_{mol} \le$\,$3.75 \times 10^4$\,\cmthree, values which are not supported by observations of any of the gas phases.

\section{Implications for high-redshift galaxies}
\label{sec:discussion}

The strong dependence of grain growth and the assembly of dust mass on ISM density has profound
implications for early galaxy evolution.
To date, only one star-forming galaxy at $z > 6$, A1689-zD1 with $z\sim7.5$, has a clear detection 
of dust emission \citep{watson15}.
This galaxy has a (lensing-corrected) stellar mass $\rm M_{\rm star}\sim 1.7\times10^9$\,\msun, and
SFR$\sim 3-9$\,\msunyr.
With an estimated dust mass of $4\times10^7$\,\msun,
the dust-to-stellar mass ratio for this galaxy is relatively high, $\sim$0.02, comparable
to the highest values found for local star-forming galaxies at similar masses \citep[e.g.,][]{skibba11}.
The age of A1689-zD1 is estimated to be $\sim$80\,Myr, at a redshift when the universe
was $\la$500\,Myr old. 

Despite extensive efforts, dust emission in other star-forming galaxies at comparable redshifts ($z\ga7$) 
has not yet been detected
\citep[e.g., IOK-1, z8-GND-5296:][]{ota14,schaerer15}.
Thus,
we predict that A1689-zD1 will be found to have a dense ISM, and thus able to assemble a considerable
dust mass over the short times available at these redshifts. 
Our findings suggest that dust mass can be a sensitive indicator of the physical conditions
in the ISM, and that metallicity alone is insufficient to determine the amount of dust.

\section{Conclusions}
\label{sec:conclusions}

In this paper we have investigated the origin of the observed dust masses in the two most metal-poor local  dwarf galaxies,
\sbs\ and \izw. Despite their comparable metallicities, gas and stellar masses, these two galaxies show a huge
variation in their dust content, with a dust mass of $\rm 3.8 \times 10^4$\,\msun\ in \sbs, as inferred by
recent ALMA observations (H14), 
and a dust mass of only 340\,\msun\ in \izw. By means of a
chemical evolution model with dust, we find that:
\begin{itemize}
\item The observed stellar population in \sbs\ can not account for the existing metal and dust masses, hence
previous stellar populations must have pre-enriched the ISM of the galaxy.
\item Even assuming the maximum possible stellar dust yields, the same stars which have pre-enriched the ISM
of \sbs\ could have injected a dust mass which is at most  15\% of the observed value. Hence, a major fraction
of the observed dust mass must originate {\it in situ} through grain growth. 
\item The observed gas density of \sbs\ is large enough to activate efficient grain growth. If 
\meannhtwo\,=\ 1500\,\cmthree,
grain growth can account for more than 85\% of the existing dust mass, with dust pre-enrichment making up the rest.
\item Despite the longer age spread estimated for the stellar population in \izw, only between 4 and 60\% of the observed
metals are produced by self-enrichment through SN explosions of massive stellar progenitors ($m > 13 - 34$\,\msun).
Hence, the metallicity has been mostly inherited by previous stellar generations. 
\item Due to the smaller gas density of \izw, grain growth is very inefficient. Since dust injected by newly-formed stars 
is partly compensated by grain destruction by interstellar shocks and astration, more than $80\%$ of the existing dust mass 
is inherited from pre-enrichment, with grain growth making up the rest.
\end{itemize}

Since dust grains can be efficiently destroyed by interstellar shocks, it is very hard to predict the inital mass of dust
that can be assembled by pre-enrichment. For this reason, we have also explored a limiting case where the galaxies
start with a negligible dust content and achieve all of their dust mass by means of grain growth. This requires the
gas density to be larger than inferred from observations: 
$5\times 10^3 \, {\rm cm}^{-3} \leq n_{\rm mol} \leq 1 \times 10^4$\,\cmthree\ for \sbs\ and 
$9.5 \times 10^3$\,\cmthree\,$\le  n_{\rm mol}\le\, 3.75 \times 10^4$\,\cmthree\ for \izw.
While  for \sbs \, $n_{\rm mol} = 5 \times 10^3 {\rm cm}^{-3}$ is within the values estimated by 
Johnson et al. (2009) for the ionized gas densities (see their Table 4), 
no observations for \izw \, suggest similarly high gas densities.
However, independently of these considerations, our study suggests that the widely different dust masses in 
\sbs\ and \izw\ reflect the different efficiencies of grain growth in their ISM, which -- given the
comparable metallicity -- is likely to originate from the different gas densities of the two galaxies.

\section*{Acknowledgments}
We thank Luca Graziani for his insightful comments and Robert Nikutta for a careful analysis of the SED fitting.
The research leading to these results has received funding from the European Research Council under the European 
Union’s Seventh Framework Programme (FP/2007-2013) / ERC Grant Agreement n. 306476.
LH is grateful to funding from INAF-PRIN 2012/2015.

\label{lastpage}


\begin{thebibliography}{99}
\bibitem[Adamo et al.(2010)]{adamo10} Adamo, A., Zackrisson, 
E., {\"O}stlin, G., \& Hayes, M.\ 2010, \apj, 725, 1620 


\bibitem[Anders 
\& Grevesse(1989)]{anders89} Anders, E., \& Grevesse, N.\ 1989, \gca, 53, 197 

\bibitem[Asano et 
al.(2013)]{asano13} Asano, R.~S., Takeuchi, T.~T., Hirashita, H., \& Inoue, A.~K.\ 2013, Earth, Planets, and Space, 65, 213 

\bibitem[Asplund et 
al.(2009)]{asplund09} Asplund, M., Grevesse, N., Sauval, A.~J., \& Scott, P.\ 2009, \araa, 47, 481 

\bibitem[Bianchi \& Schneider(2007)]{bianchi07} Bianchi, S., \& Schneider, R.\ 2007, \mnras, 378, 973 

\bibitem[Bocchio et 
al.(2014)]{bocchio14} Bocchio, M., Jones, A.~P., \& Slavin, J.~D.\ 2014, \aap, 570, AA32 

\bibitem[Cannon et al.(2002)]{cannon02} Cannon, J.~M., Skillman, 
E.~D., Garnett, D.~R., \& Dufour, R.~J.\ 2002, \apj, 565, 931 

\bibitem[Cannon et al.(2005)]{cannon05} Cannon, J.~M., Walter, 
F., Skillman, E.~D., \& van Zee, L.\ 2005, \apjl, 621, L21 

\bibitem[Cormier et 
al.(2012)]{cormier12} Cormier, D., Lebouteiller, V., Madden, S.~C., et al.\ 2012, \aap, 548, A20 

\bibitem[de Bennassuti et al.(2014)]{debennassuti14} de Bennassuti, 
M., Schneider, R., Valiante, R., \& Salvadori, S.\ 2014, \mnras, 445, 3039 

\bibitem[Draine et al.(2007)]{draine07sings} Draine, B.~T., et al.\ 
2007, \apj, 663, 866 

\bibitem[Draine 
\& Li(2007)]{draine07} Draine, B.~T., \& Li, A.\ 2007, \apj, 657, 810 

\bibitem[Dwek(1998)]{dwek98} Dwek, E.\ 1998, \apj, 501, 643 

\bibitem[Ekta et al.(2009)]{ekta09} Ekta, B., Pustilnik, 
S.~A., \& Chengalur, J.~N.\ 2009, \mnras, 397, 963 

\bibitem[Fisher et al.(2014)]{fisher14} Fisher, D.~B., Bolatto, 
A.~D., Herrera-Camus, R., et al.\ 2014, \nat, 505, 186 

\bibitem[Fumagalli et al.(2010)]{fumagalli10} Fumagalli, M., 
Krumholz, M.~R., \& Hunt, L.~K.\ 2010, \apj, 722, 919 

\bibitem[Galametz et 
al.(2011)]{galametz11} Galametz, M., Madden, S.~C., Galliano, F., et al.\ 2011, \aap, 532, A56 

\bibitem[Giammanco et 
al.(2004)]{giammanco04} Giammanco, C., Beckman, J.~E., Zurita, A., \& Rela{\~n}o, M.\ 2004, \aap, 424, 877 

\bibitem[Gnedin et al.(2009)]{gnedin09} Gnedin, N.~Y., Tassis, 
K., \& Kravtsov, A.~V.\ 2009, \apj, 697, 55 

\bibitem[Granato et al.(2000)]{granato00} Granato, G.~L., Lacey, 
C.~G., Silva, L., et al.\ 2000, \apj, 542, 710 

\bibitem[Hirashita(2015)]{hirashita15} Hirashita, H.\ 2015, \mnras, 447, 2937 

\bibitem[Hirashita et al.(2014)]{hirashita14} Hirashita, H., 
Ferrara, A., Dayal, P., \& Ouchi, M.\ 2014, \mnras, 443, 1704 

\bibitem[Hirashita \& Voshchinnikov(2014)]{hirashitavosh14} Hirashita, H., \& Voshchinnikov, N.~V.\ 2014, \mnras, 437, 1636 

\bibitem[Houck et al.(2004)]{houck04} Houck, J.~R., 
Charmandaris, V., Brandl, B.~R., et al.\ 2004, \apjs, 154, 211 

\bibitem[Hunt et 
al.(2005a)]{hunt05a} Hunt, L., Bianchi, S., \& Maiolino, R.\ 2005a, \aap, 434, 849 

\bibitem[Hunt et 
al.(2005b)]{hunt05b} Hunt, L.~K., Dyer, K.~K., \& Thuan, T.~X.\ 2005b, \aap, 436, 837 

\bibitem[Hunt et al.(2004)]{hunt04} Hunt, L.~K., Dyer, K.~K., 
Thuan, T.~X., \& Ulvestad, J.~S.\ 2004, \apj, 606, 853 

\bibitem[Hunt 
\& Hirashita(2009)]{hunt09} Hunt, L.~K., \& Hirashita, H.\ 2009, \aap, 507, 1327 

\bibitem[Hunt et al.(2012)]{hunt12} Hunt, L., Magrini, L., 
Galli, D., et al.\ 2012, \mnras, 427, 906 

\bibitem[Hunt et al.(2014)]{hunt14} Hunt, L.~K., Testi, L., Casasola, V., et al.\ 2014, \aap, 561, AA49 

\bibitem[Hunt et al.(2003)]{hunt03} Hunt, L.~K., Thuan, T.~X., 
\& Izotov, Y.~I.\ 2003, \apj, 588, 281 

\bibitem[Hunt et 
al.(2001)]{hunt01} Hunt, L.~K., Vanzi, L., \& Thuan, T.~X.\ 2001, \aap, 377, 66 

\bibitem[Issa et 
al.(1990)]{issa90} Issa, M.~R., MacLaren, I., \& Wolfendale, A.~W.\ 1990, \aap, 236, 237 

\bibitem[Ivezic 
\& Elitzur(1997)]{ivezic97} Ivezic, Z., \& Elitzur, M.\ 1997, \mnras, 287, 799 

\bibitem[Izotov et al.(1999)]{izotov99} Izotov, Y.~I., Chaffee, 
F.~H., Foltz, C.~B., et al.\ 1999, \apj, 527, 757 

\bibitem[Izotov et 
al.(2014)]{izotov14} Izotov, Y.~I., Guseva, N.~G., Fricke, K.~J., Kr{\"u}gel, E., \& Henkel, C.\ 2014, \aap, 570, A97 

\bibitem[Johnson et al.(2009)]{johnson09} Johnson, K.~E., Hunt, 
L.~K., \& Reines, A.~E.\ 2009, \aj, 137, 3788 

\bibitem[Jones et al. (2013)]{jones13} 
Jones A.~P.,  Fanciullo L., K{\"o}hler M., Verstraete L., Guillet V., Bocchio M., Ysard N. 2013, A\&A, 558, 62

\bibitem[Jones \& Nuth(2011)]{jones11} Jones, A.~P., \& Nuth, J.~A.\ 2011, \aap, 530, AA44 

\bibitem[Kassim et al.(1989)]{kassim89} Kassim, N.~E., Weiler, 
K.~W., Erickson, W.~C., \& Wilson, T.~L.\ 1989, \apj, 338, 152 

\bibitem[Kennicutt(1984)]{kennicutt84} Kennicutt, R.~C., Jr.\ 1984, 
\apj, 287, 116 

\bibitem[Krumholz et al.(2009a)]{krumholz09a} Krumholz, M.~R., 
McKee, C.~F., \& Tumlinson, J.\ 2009a, \apj, 693, 216 

\bibitem[Krumholz et al.(2009b)]{krumholz09b} Krumholz, M.~R., 
McKee, C.~F., \& Tumlinson, J.\ 2009b, \apj, 699, 850 

\bibitem[Krumholz et al.(2012)]{krumholz12} Krumholz, M.~R., Dekel, A., \& McKee, C.~F.\ 2012, \apj, 745, 69 

\bibitem[Lebouteiller et 
al.(2012)]{lebout12} Lebouteiller, V., Cormier, D., Madden, S.~C., et al.\ 2012, \aap, 548, A91 

\bibitem[Lelli et 
al.(2012)]{lelli12} Lelli, F., Verheijen, M., Fraternali, F., \& Sancisi, R.\ 2012, \aap, 537, AA72 

\bibitem[Leroy et al.(2011)]{leroy11} Leroy, A.~K., Bolatto, 
A., Gordon, K., et al.\ 2011, \apj, 737, 12 


\bibitem[Madden et al.(2013)]{madden13} Madden, S.~C., 
R{\'e}my-Ruyer, A., Galametz, M., et al.\ 2013, \pasp, 125, 600 

\bibitem[Magrini et 
al.(2011)]{magrini11} Magrini, L., Bianchi, S., Corbelli, E., et al.\ 2011, \aap, 535, AA13 

\bibitem[Martin(1997)]{martin97} Martin, C.~L.\ 1997, \apj, 491, 
561 

\bibitem[Mu{\~n}oz-Mateos et al.(2009)]{munoz09} 
Mu{\~n}oz-Mateos, J.~C., Gil de Paz, A., Boissier, S., et al.\ 2009, \apj, 
701, 1965 

\bibitem[Ota et al.(2014)]{ota14} Ota, K., Walter, F., Ohta, 
K., et al.\ 2014, \apj, 792, 34 

\bibitem[Peeples et al.(2014)]{peeples14} Peeples, M.~S., Werk, 
J.~K., Tumlinson, J., et al.\ 2014, \apj, 786, 54 

\bibitem[Plante 
\& Sauvage(2002)]{plante02} Plante, S., \& Sauvage, M.\ 2002, \aj, 124, 1995 

\bibitem[Pustilnik et al.(2001)]{pustilnik01} Pustilnik, S.~A., 
Brinks, E., Thuan, T.~X., Lipovetsky, V.~A., 
\& Izotov, Y.~I.\ 2001, \aj, 121, 1413 

\bibitem[Raiteri et al.(1996)]{raiteri96} Raiteri, C.~M., Villata, M., \& Navarro, J.~F.\ 1996, \aap, 315, 105 

\bibitem[Recchi et 
al.(2002)]{recchi02} Recchi, S., Matteucci, F., D'Ercole, A., \& Tosi, M.\ 2002, \aap, 384, 799 

\bibitem[Reines et al.(2008)]{reines08} Reines, A.~E., Johnson, 
K.~E., \& Hunt, L.~K.\ 2008, \aj, 136, 1415 

\bibitem[Reines et al.(2010)]{reines10} Reines, A.~E., Nidever, 
D.~L., Whelan, D.~G., \& Johnson, K.~E.\ 2010, \apj, 708, 26 

\bibitem[R{\'e}my-Ruyer et 
al.(2013)]{remy13} R{\'e}my-Ruyer, A., Madden, S.~C., Galliano, F., et al.\ 2013, \aap, 557, A95 

\bibitem[R{\'e}my-Ruyer et 
al.(2014)]{remy14} R{\'e}my-Ruyer, A., Madden, S.~C., Galliano, F., et al.\ 2014, \aap, 563, AA31 

\bibitem[Salvadori et al.(2007)]{salvadori07} Salvadori, S., 
Schneider, R., \& Ferrara, A.\ 2007, \mnras, 381, 647 

\bibitem[Schaerer et 
al.(2015)]{schaerer15} Schaerer, D., Boone, F., Zamojski, M., et al.\ 2015, \aap, 574, AA19 

\bibitem[Schmidt 
\& Boller(1993)]{schmidt93} Schmidt, K.-H., \& Boller, T.\ 1993, Astronomische Nachrichten, 314, 361 

\bibitem[Skibba et al.(2011)]{skibba11} Skibba, R.~A., 
Engelbracht, C.~W., Dale, D., et al.\ 2011, \apj, 738, 89 

\bibitem[Smith \& Hancock(2009)]{smith09} Smith, B.~J., \& Hancock, M.\ 2009, \aj, 138, 130 

\bibitem[Thuan \& Izotov(1997)]{thuan97} Thuan, T.~X., \& Izotov, Y.~I.\ 1997, \apj, 489, 623 

\bibitem[Thuan et al.(1997)]{thuan97ssc} Thuan, T.~X., Izotov, 
Y.~I., \& Lipovetsky, V.~A.\ 1997, \apj, 477, 661 

\bibitem[Thuan et al.(1999)]{thuan99} Thuan, T.~X., Sauvage, 
M., \& Madden, S.\ 1999, \apj, 516, 783 

\bibitem[Valiante et al.(2009)]{valiante09} Valiante, R., 
Schneider, R., Bianchi, S., \& Andersen, A.~C.\ 2009, \mnras, 397, 1661 

\bibitem[Valiante et al.(2011)]{valiante11} Valiante, R., 
Schneider, R., Salvadori, S., \& Bianchi, S.\ 2011, \mnras, 416, 1916 

\bibitem[Valiante et al.(2012)]{valiante12} Valiante, R., 
Schneider, R., Maiolino, R., Salvadori, S., \& Bianchi, S.\ 2012, \mnras, 427, L60 

\bibitem[Valiante et al.(2014)]{valiante14} Valiante, R., 
Schneider, R., Salvadori, S., \& Gallerani, S.\ 2014, \mnras, 444, 2442 

\bibitem[van den Hoek \& Groenewegen(1997)]{vandenhoek97} van den Hoek, L.~B., \& Groenewegen, M.~A.~T.\ 1997, \aaps, 123, 305 

\bibitem[van Zee et al.(1998)]{vanzee98} van Zee, L., Westpfahl, 
D., Haynes, M.~P., \& Salzer, J.~J.\ 1998, \aj, 115, 1000 

\bibitem[Vanzi et 
al.(2000)]{vanzi00} Vanzi, L., Hunt, L.~K., Thuan, T.~X., \& Izotov, Y.~I.\ 2000, \aap, 363, 493 

\bibitem[Watson et al.(2015)]{watson15} Watson, D., Christensen, 
L., Kraiberg Knudsen, K., et al.\ 2015, arXiv:1503.00002, \nat, in press 

\bibitem[Wolfire et al.(2003)]{wolfire03} Wolfire, M.~G., McKee, 
C.~F., Hollenbach, D., \& Tielens, A.~G.~G.~M.\ 2003, \apj, 587, 278 

\bibitem[Woosley \& Weaver(1995)]{woosley95} Woosley, S.~E., \& Weaver, T.~A.\ 1995, \apjs, 101, 181 

\bibitem[Zaritsky et al.(1994)]{zaritsky94} Zaritsky, D., 
Kennicutt, R.~C., Jr., \& Huchra, J.~P.\ 1994, \apj, 420, 87 

\bibitem[Zhukovska et al.(2008)]{zhukovska08} Zhukovska, S., Gail, H.-P., \& Trieloff, M.\ 2008, \aap, 479, 453 
\end{thebibliography}
\end{document}